\documentclass[fleqn,10pt]{wlscirep}
\usepackage[utf8]{inputenc}
\usepackage[T1]{fontenc}
\usepackage{bm}
\usepackage[figurename=Fig.]{caption}

\title{Experimental demonstration of crowd synchrony and first-order transition with lasers} 

\author[1,*]{Simon Mahler}
\author[1]{Asher A. Friesem}
\author[1]{Nir Davidson}
\affil[1]{Department of Physics of Complex Systems, Weizmann Institute of Science, Rehovot 761001, Israel}
\affil[*]{e-mail: sim.mahler@gmail.com}

\begin{abstract}
\textbf{Crowd synchrony, which corresponds to the synchronization of different and independent oscillators that interact with each other via a common intermediate, is ubiquitous in many fields. Here, we experimentally demonstrate the effect of crowd synchrony, analogous to that of the Millennium Bridge, by resorting to coupled lasers. When the number of lasers is below a critical number, there is no synchronization, but after reaching the critical number, the lasers instantaneously synchronize. We show that the synchronization of the lasers as a function of their number follows a first-order-like transition, and that our experimental results are in good agreement with those predicted by theoretical models.}
\end{abstract}

\begin{document}
\flushbottom
\maketitle
\thispagestyle{empty}

\section*{Introduction}
Synchronization of independent dynamical elements via an intermediate medium plays an important role in biology, chemistry, engineering and physics. It was investigated in cell and molecular biology \cite{Aldridge76,DeMonte07,Konigsberg71,Dano99,Camilli06}, chemical oscillators \cite{Taylor09,Toth06,Tinsley10}, bridge engineering \cite{Dallard01,Strogatz05,Eckhardt07,Belykh17,Ingolfsson11,Zivanovic05}, humans walking \cite{Joshi18,Ingolfsson11}, optics and lasers \cite{Zamora10,Cohen12}, collective behavior of species and microorganisms \cite{Gregor10}, clocks \cite{Danino10,Garcia-Ojalvo04}, chaotic oscillators \cite{Gomez-Gardenes11,Leyva12,Singh15} and networks \cite{Gomez-Gardenes11,Leyva12,Gambuzza15}. In most of the cases, the synchronization depends on the number of elements where a critical number determines the onset of the synchronization. Representative examples where such an onset occurs are in quorum sensing and crowd synchrony. 
In quorum sensing, which was initially observed when studying collective behaviors and interactions of cells \cite{Konigsberg71,Martz72,Aldridge76}, the dynamics and oscillations of the elements only occurs above a critical density with a first or second-order-like synchronization transition \cite{Taylor09,Zamora10,Singh15,Aldridge76}. 

Crowd synchrony was first investigated and modeled on the pedestrian Millennium Bridge in London, which was closed shortly after its opening day due to uncontrolled strong lateral oscillations. The pedestrians, each walking at different pace and speed, caused small lateral oscillations to the bridge, which in turn caused the pedestrians to sway in step in order to retain balance, dramatically amplifying the oscillations of the bridge and synchronizing the pedestrians \cite{Dallard01,Joshi18,Eckhardt07,Zivanovic05}. The effect was modeled as crowd synchronization (Fig.~\ref{fig:1_exp_skecth}a), where the lateral oscillations of the bridge were attributed to a combination of oscillations and synchronization that critically depend on the number of pedestrians \cite{Strogatz05,Eckhardt07}. 

Crowd synchrony was theoretically predicted also with coupled lasers, where coupling between $M$ independent  (star) lasers (analogous to pedestrians) is mediated by a central (hub) laser (analogous to the bridge) operating below lasing threshold (Fig.~\ref{fig:1_exp_skecth}b) \cite{Zamora10}. When $M$ is above a critical number $M_{c}$, the lasers synchronize with a first-order-like transition where $M_{c}$ depends on the coupling strength between the star lasers and the hub laser, on the frequency distribution of the lasers and on the pumping strength \cite{Zamora10,Cohen12,Eckhardt07}.

In this work, we present a first experimental demonstration of crowd synchrony with coupled lasers. It is achieved with two coupled degenerate cavity lasers (DCLs). The first DCL forms a controllable number of independent (star) lasers with nearly uniform intensities but different phases and frequencies. The second (hub) DCL, operating below lasing threshold, can phase lock (synchronize) the independent star lasers when their number increase. As in crowd synchrony, we observed a first-order-like transition to synchronization as the number of star lasers crosses a critical number $M_{c}$ that depends on the coupling strength $K$ between the two DCLs and follows a power law with a scaling exponent $\nu=-2.4$, in good agreement with the theoretically predicted one $\nu_{th}=-2.2$ \cite{Zamora10}. This scaling law is independent of the array geometry, providing evidence of universality. We also show that the sharp transition to synchronization is related to the onset of the lasing transition of the hub laser, providing the feedback needed for a first-order-like transition \cite{Taylor09}.

\section*{Experimental arrangement and results}
\begin{figure}[!ht]
\centering
\includegraphics[width=0.48\textwidth]{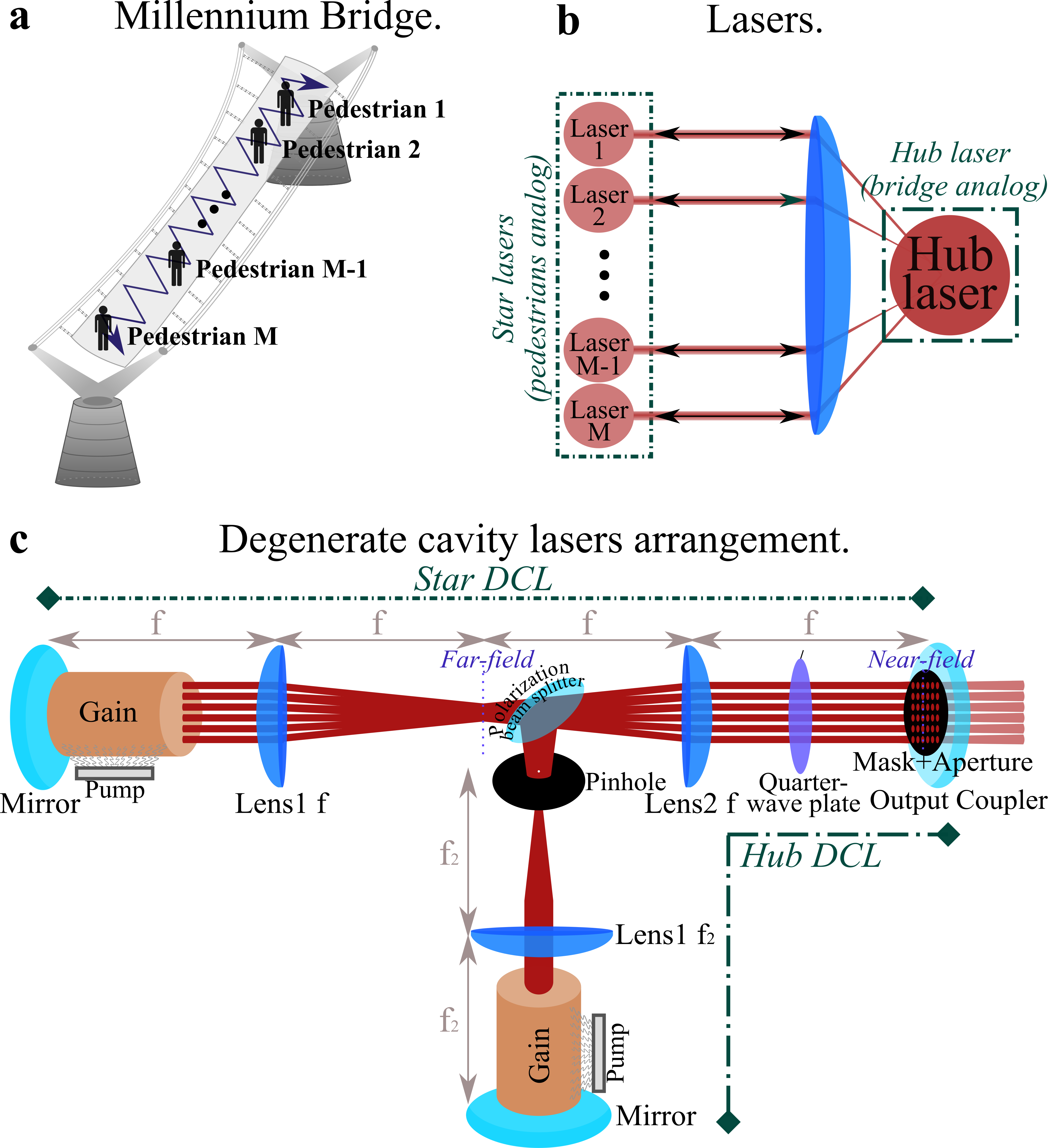}
\caption{\textbf{Crowd synchronization.} \textbf{a}, Millennium Bridge, which is immobile with few pedestrians and laterally oscillating with many pedestrians \cite{Dallard01,Strogatz05}. \textbf{b}, Model for crowd synchrony with lasers, where there is no phase locking (synchronization) with few lasers and phase locking with many lasers\cite{Zamora10}. \textbf{c}, Arrangement for experimentally demonstrating crowd synchrony composed of two self-imaging degenerate cavity lasers (DCLs) coupled with adjustable coupling strength $K$. The multi-mode DCL serves to form independent (uncoupled) star lasers and the hub laser DCL provides long range coupling between them. A variable size near-field aperture and an amplitude mask adjacent to the output coupler control the number of star lasers.}
\label{fig:1_exp_skecth}
\end{figure}

Our experimental arrangement for investigating crowd synchrony with coupled lasers is schematically presented in Fig.~\ref{fig:1_exp_skecth}c. It includes two DCLs coupled with a tunable coupling strength $K$. One DCL (horizontal orientation) has a high number of transverse modes and forms at the near-field plane an array of independent and uncoupled (star) lasers with any desired geometry \cite{Tradonsky17,Mahler19}. A variable size aperture controls the number of star lasers. A second (hub) DCL with a single transverse mode mediates positive and long-range coupling between the star lasers \cite{Tradonsky17}. The intensity distributions at the near-field plane of the star lasers array and at its far-field plane (corresponding to its coherence function) are both imaged to and detected by a CCD camera. The star lasers operates above lasing threshold and the hub laser below lasing threshold. More details are given in the Methods. 

\begin{figure}[!ht]
\centering
\includegraphics[width=0.75\textwidth]{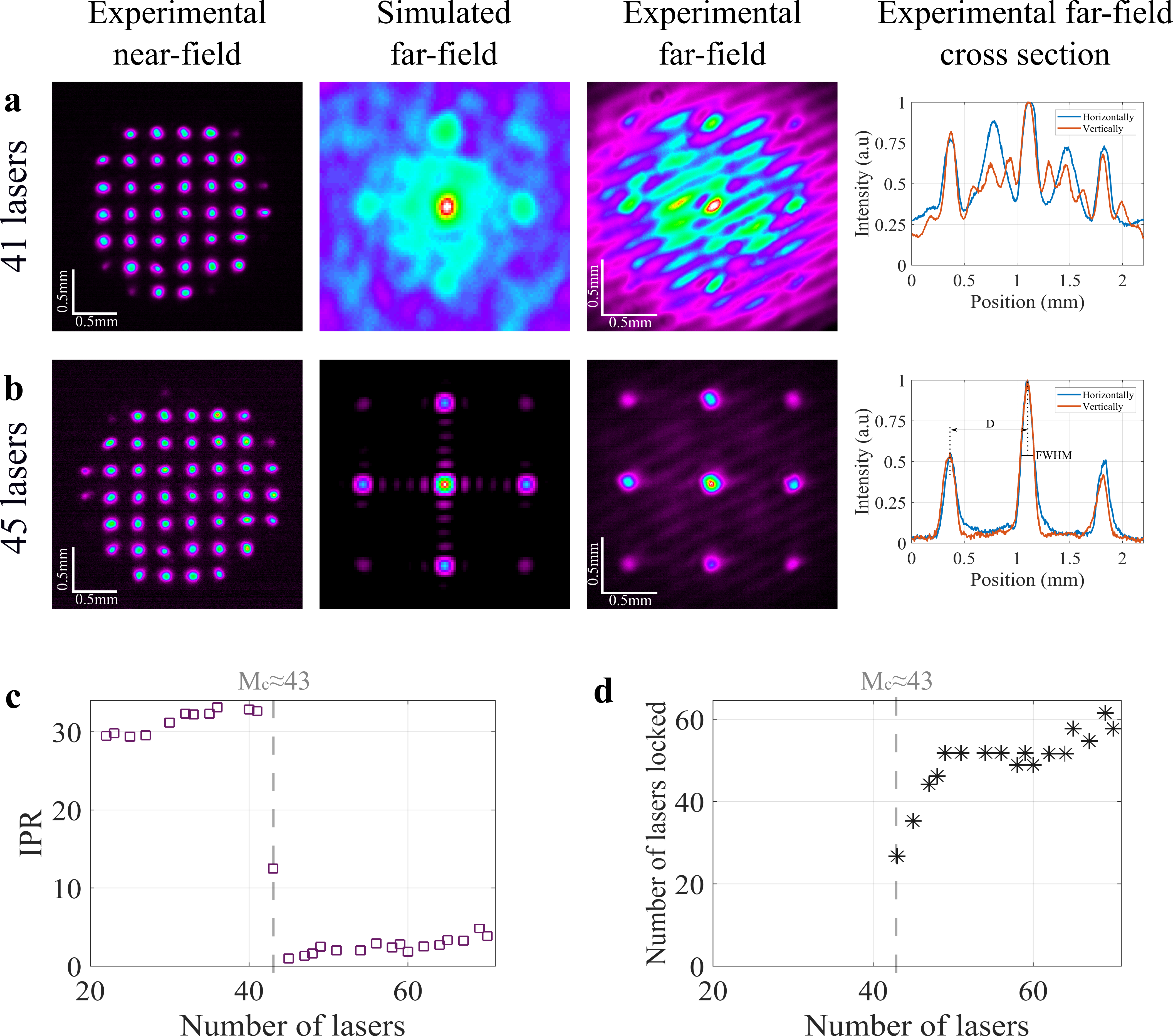}
\caption{\textbf{Crowd synchronization in a square array of lasers.} \textbf{a and b}, Experimental and simulated near-field and far-field intensity distributions with corresponding cross sections along the far-field center for square array of star lasers with $log(K)\approx3.6$. \textbf{a}, $M=41$ lasers where the broad far-field intensity distribution indicates no phase synchronization. \textbf{b}, $M=45$ lasers, where the sharp far-field peaks indicate phase synchronization. \textbf{c}, Inverse participation ratio (IPR) as a function of the number of star lasers revealing a sharp transition to synchronization, at a critical number $M_{c}=43$. \textbf{d}, Number of phase locked (synchronized) lasers as a function of the number of star lasers.}
\label{fig:2_crwdsynch_square}
\end{figure}

\begin{figure}[!ht]
\centering
\includegraphics[width=0.75\textwidth]{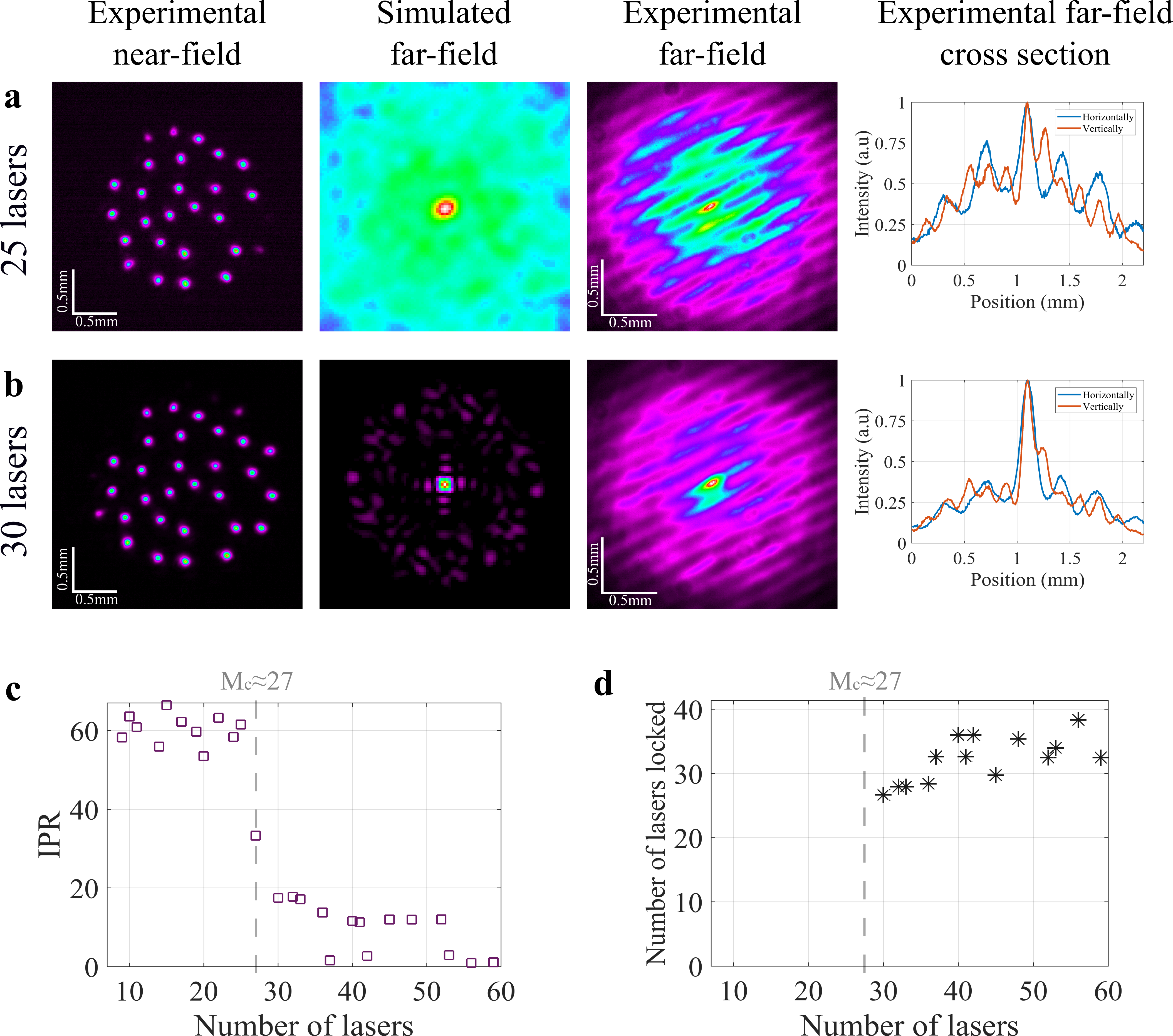}
\caption{\textbf{Crowd synchronization in a random array of lasers.} \textbf{a and b}, Experimental and simulated near-field and far-field intensity distributions with corresponding cross section intensities along the far-field center for random array of star lasers with $log(K)\approx3.8$. \textbf{a}, $M=25$ lasers where the broad far-field intensity distribution indicates little or no phase synchronization and \textbf{b}, $M=30$ lasers where the sharp far-field intensity peak indicates phase locking (synchronization). \textbf{c}, Inverse participation ratio (IPR) as a function of the number of star lasers revealing a sharp transition to synchronization, at a critical number $M_{c}=27$. \textbf{d}, Number of phase locked (synchronized) lasers as a function of the number of star lasers.}
\label{fig:3_crwdsynch_random}
\end{figure}

\begin{figure}[!ht]
\centering
\includegraphics[width=0.4 \textwidth]{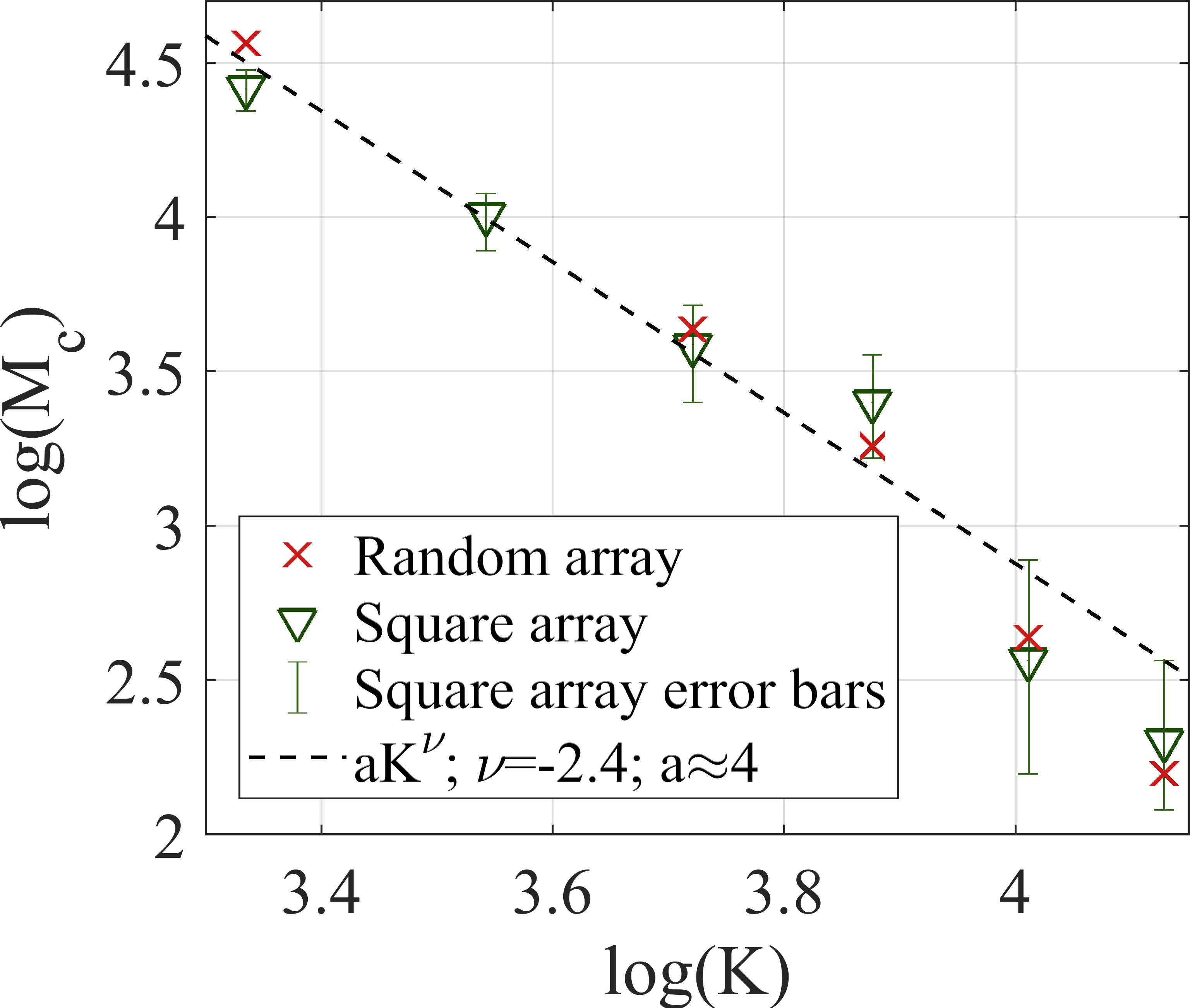}
\caption{\textbf{Universality and critical exponent.} Critical number of lasers $M_{c}$ as a function of the coupling strength $K$ between the two DCLs for square and random laser arrays. The critical number of lasers decreases as the coupling increases, and is well fitted by a power law $M_{c}\propto K^{\nu}$ with $\nu=-2.4\pm0.35$ (dashed line). Error bars for the random array (not shown) are similar to those of the square array.}
\label{fig:4_M_c_vs_K}
\end{figure}

Using the arrangement of Fig.~\ref{fig:1_exp_skecth}c, we performed extensive experiments and simulations with square and with random arrays of lasers to demonstrate and study crowd synchronization. The results are presented in Figs.~\ref{fig:2_crwdsynch_square}-\ref{fig:4_M_c_vs_K}. Figure~\ref{fig:2_crwdsynch_square}a and \ref{fig:2_crwdsynch_square}b show the near-field and far-field intensity distributions and their cross sections, for two slightly different numbers of lasers in a square array. As evident, the near-field intensity distributions for the two arrays are similar, whereas their far-field intensity distributions dramatically differ. For $M=41$ lasers (Fig.~\ref{fig:2_crwdsynch_square}a, slightly below $M_{c}$), the far-field intensity distribution and its cross-section reveal a broad Gaussian envelope with some weak internal fringes, indicating that the lasers are uncoupled \cite{Mahler19,Tradonsky17}. For $M=45$ lasers (Fig.~\ref{fig:2_crwdsynch_square}b, slightly above $M_{c}$), the far-field intensity distribution and it cross-section contain distinct sharp peaks indicating stable in-phase locking (synchronization) between the lasers. We determined that $80\%$ of the $M=45$ lasers were phase synchronized, see Methods. Numerical simulations performed by a combined algorithm (see Methods) are in good agreement with the experimental data. 

Both the experimental and simulation results indicate that synchronization critically depends on the number of star lasers. Below a critical number ($M_{c}=43$ lasers here), the star lasers remain uncoupled whereby their energy is not sufficient to pump and enforce lasing in the hub laser. Such behavior is analogous to that occurring in the Millennium Bridge where below a critical number of pedestrians there are no lateral oscillations. Above $M_{c}$, the star lasers phase synchronize thereby providing sufficiently strong coherent signal to drive the hub laser above its lasing transition, where it couples the star lasers and synchronize them. 

To quantify the results, we repeated these experiments for many different numbers of star lasers and characterized the synchronization by the inverse participation ratio (IPR) of the measured far-field intensity distributions: 
\begin{equation}
IPR=\frac{\left[\sum_{i}I_{i}\right]^{2}}{\sum_{i}I_{i}^{2}},
\label{eq:1_IPR}
\end{equation}
where $I_{i}$ is the intensity of the $i$th pixel in the far-field and the sums occur over all the pixels. The IPR indicates whether sharp peaks occur in the far-field intensity distribution and is therefore a quantitative measure of synchronization\cite{Mahler19,Tradonsky17}. Figure.~\ref{fig:2_crwdsynch_square}c shows the IPR as a function of the number of star lasers. As evident, a sharp transition occurs at $M_{c}=43$ between high and low plateau values of the IPR. The low IPR plateau above $M_{c}=43$ correspond to sharp intensity peaks, as would occurs for phase synchronized lasers. Figure~\ref{fig:2_crwdsynch_square}d shows the number of lasers that are phase synchronized (locked) (see Methods and Supplementary Information S2B) as a function of the number of lasers. At $M>M_{c}$, the number of phase locked lasers sharply increases and continues to increase, indicating that most (above $80\%$) of the lasers are synchronized.

Next, we repeated all the experiments and measurements for a random array of lasers where each laser was randomly positioned in space with the condition that the distance to its nearest neighbors was $a=300$ $\mu$m. The results are presented in Fig.~\ref{fig:3_crwdsynch_random}. Figures~\ref{fig:3_crwdsynch_random}a and \ref{fig:3_crwdsynch_random}b show representative experimental and simulated near-field and far-field intensity distributions for $M=25$ and $M=30$ lasers, respectively. For $M=25$ lasers, the far-field intensity distribution is broadly distributed, indicating that the lasers are uncoupled \cite{Mahler19,Tradonsky17}. For $M=30$ lasers, the far-field intensity distribution contains a sharp central peak, indicating that most of the lasers are phase synchronized \cite{Mahler19,Tradonsky17}. 

Figure~\ref{fig:3_crwdsynch_random}c shows the IPR of the far-field intensity distributions as a function of the number of lasers. As for the square array, a sharp first-order-like synchronization transition occurs at a critical number $M_{c}=27$ lasers. Figure~\ref{fig:3_crwdsynch_random}d shows that above $M_{c}$ most of the lasers are phase synchronized. The results for the random laser array are similar to those of the square array, indicating that the geometry of the array does not influence the synchronization of the lasers.

We repeated the experiments and measurements for both square and random laser arrays with different coupling strengths $K$ between the star and the hub DCLs, and determined the critical number of lasers $M_{c}$ as a function of $K$. The results, presented in Fig.~\ref{fig:4_M_c_vs_K} indicate that $M_{c}$ does not depend on the array geometry and decreases as the coupling increases. The results for both array geometries are well fitted by a power law $M_{c}\propto K^{\nu}$ with a scaling exponent $\nu=-2.4$, in agreement with the theoretically predicted one $\nu_{th}=-2.2$ for time-delayed coupled lasers \cite{Zamora10} but significantly different from the scaling exponent $\nu_{th}=-1$ predicted for pedestrians walking on the Millennium Bridge \cite{Strogatz05}. We attribute this difference to the highly nonlinear lasing transition of the hub laser that plays an important role in synchronizing the star lasers (as opposed to the linear response of the bridge to the force applied by the pedestrians) . 

Finally, we studied the relation between the crowd synchrony transition and the lasing transition of the hub laser. We found that the hub laser starts to lase when the number of star lasers reach $M_{c}$ and even slightly below it, similar to the dynamics of feedback-stabilized walking bipeds \cite{Joshi18}. Conversely, when the hub laser is set above lasing threshold, synchronization depends smoothly on the number of star lasers \cite{Zamora10}, in analogy to crowd dynamics on a wobbly bridge \cite{Belykh17}. 
Similar synchronization situation occurs in coupled chemical oscillators \cite{Taylor09}. For weak coupling, the auto-catalyst concentration of each oscillator is unaffected by the other oscillators that leads to a smooth transition to synchronization as the density of oscillators is increased. For strong coupling, each oscillator is affected by the other oscillators with a positive feedback that leads to a dynamical instability and a sharp transition to synchronization. Chemical oscillators with fixed concentrations and also lasers with fixed intensities are well described by the Kuramoto model, that predicts a second-order-like transition even for all to all couplings. Sharp first-order-like transition can only occur when the chemical concentrations or laser intensities can vary \cite{Kourtchatov95,Ricardo01}.     

\section*{Conclusions}
To conclude, we have shown experimentally that an array of star lasers coupled to a common hub laser can synchronize, similarly to crowd synchrony of pedestrians on the Millennium Bridge. The synchronization follows a first-order-like transition as the number of the lasers crosses a critical number. The critical number of lasers obeys a power law as a function of the coupling strength with a scaling exponent $\nu=-2.4$, independent of the array geometry but different from that of the Millennium bridge, reflecting the important role of the highly nonlinear lasing transition in synchronization. The transition to synchronization of coupled lasers by the lasing transition has attracted a lot of attention for a variety of research activities, simulating spin systems \cite{Mahler20,Tradonsky19}, solving computational problems \cite{Nixon13,McMahon16,Berloff17}, studying topological dynamics \cite{Pal17}, and coherent beam combining \cite{Mahler19,Tradonsky17}. Crowd synchrony and its associated sharp and universal transitions can advance our understanding in these research areas and also open new ones.
Our investigations were centered on crowd synchrony, but the experimental arrangement could be adapted to deal with more general synchronization or quorum sensing situations.

\section*{Methods}
\subsection*{Experimental arrangement}
The experimental arrangement schematically presented in Fig.~\ref{fig:1_exp_skecth}c includes two self-imaging DCLs \cite{Arnaud69} coupled together by means of a quarter-wave plate and a polarization beam splitter that control the coupling strength $K$ between them \cite{Nixon11}. The first DCL (horizontal orientation) has a high number of transverse modes and forms an array of independent and uncoupled lasers \cite{Tradonsky17,Mahler19}, corresponding to the star lasers. We estimated the level of frequency detuning between the star lasers to be few $MHz$. The second DCL has a single transverse mode and serves to couple the lasers positively with long range \cite{Tradonsky17}, corresponding to the hub laser. Each DCL is comprised of a back mirror of high $>95\%$ reflectivity, a Nd:YAG gain medium rod of $0.50$ cm diameter and $10.9$ cm length (operating wavelength $\lambda=1064$ nm) that is optically pumped by a quasi-CW $100$ $\mu$s duration pulsed flash lamp operating at $1$ Hz to minimize thermal lensing, a spherical lens $Lens1$ of focal length $f=20$ cm for the star lasers or $f_{2}=15$ cm for the hub laser at focal distance from the back mirror, a polarization beam splitter that combines the laser fields of the star and hub lasers, forming a common laser field, a second spherical lens $Lens2$ of focal length $f=20$ cm at focal distance $f$ from the focal plane of $Lens1$ and a common mirror to both laser fields of $95\%$ reflectivity at the front focal plane of $Lens2$ that serves as an output coupler.  

A mask of a square (or random) array of holes with diameter $200$ $\mu$m separated by a period $a=300$ $\mu$m, adjacent to the output coupler (denoted as the near-field) is used to form an array of star lasers. A variable size aperture is placed next to the mask of holes to control the number of star lasers. Lenses $Lens1$ and $Lens2$ form a telescope configuration yielding a self-imaging condition, ensuring that each hole in the mask forms an independent laser \cite{Mahler19,Tradonsky17}. At the focal plane midway between $Lens1$ and $Lens2$ of the hub laser (far-field Fourier plane), a pinhole of size below the diffraction limit couples the lasers positively with long range coupling \cite{Tradonsky17}. The near-field and far-field intensity distributions are both imaged to and detected by a CCD camera. The star lasers operate above lasing threshold and the hub laser below lasing threshold. All the far-field intensity distributions that support the plots within this Letter were averaged over five realizations.

\subsection*{Numerical simulation}
For the simulation of the far-field intensity distributions in Figs.~\ref{fig:2_crwdsynch_square} and \ref{fig:3_crwdsynch_random}, we resorted to a combined algorithm \cite{Nixon13_2} that includes the Fox-Li \cite{FoxLi61} and the Gerchberg-Saxton \cite{Gerchberg72} algorithms. It is an iterative algorithm that numerically mimics the propagation in space of the laser field in the DCL where an iteration in the algorithm numerically mimics a round-trip of the laser field in the DCL. For that, a two dimensional matrix $E_{ij}$ representing the spatial distribution of the laser field is initialized with uniform intensities and random phases. During each iteration, the initial laser field $E_{ij}$ is multiplied by a saturable gain function $G_{ij}$, then is Fourier transformed (first  far-field $FF1_{ij}$), then is inversely Fourier transformed (near-field $NF_{ij}$), then is Fourier transformed again (second far-field $FF2_{ij}$) and finally is inversely Fourier transformed (laser field $E_{ij}$). The resulting laser field $E_{ij}$ is kept and used for the next iteration. 

The saturable gain function is $G_{ij}=\frac{G_{0}}{1+|E_{ij}|^{2}/I_{sat}}$ where $G_{0}$ is the constant pump value and $I_{sat}$ the saturation intensity. Components inserted in the near-field or far-field planes of the DCL, such as mask of holes or pinhole, can be numerically mimicked in the algorithm by multiplying the near-field $NF_{ij}$ or far-field $FF_{ij}$ matrices by the $Pinhole_{ij}$ or $Mask_{ij}$ matrices. 

The experimental arrangement of Fig.~\ref{fig:1_exp_skecth}(c) was simulated by using two different laser fields $E_{ij}^{star}$ and $E_{ij}^{hub}$ each one representing the star and hub DCLs respectively. For the star DCL, the laser field $E_{ij}^{star}$ is multiplied by $G_{ij}^{star}$, Fourier transformed, inversely Fourier transformed, multiplied by $Mask_{ij}$, phase detuned (by adding random phases with $1\%$ intensity to the field), Fourier transformed, added to the laser field from the hub laser with coupling $1-K$, and finally inversely Fourier transformed. For the hub DCL, the laser field $E_{ij}^{hub}$ is multiplied by $G_{ij}^{hub}$, Fourier transformed, multiplied by $Pinhole_{ij}$, inversely Fourier transformed, multiplied by $Mask_{ij}$, Fourier transformed, added to the laser field from the star laser with coupling $K$, and finally inversely Fourier transformed. The pump value $G_{0}^{star}$ of the star DCL was set above lasing threshold and the pump value $G_{0}^{hub}$ of the hub DCL was set below lasing threshold. The mask of holes was either a square or random array geometry and the pinhole size was set below the diffraction limit of the DCL. The simulations were averaged over $100$ different realizations to account for the different longitudinal modes of the DCL \cite{Nixon13_2}. 

\subsection*{Calculating IPR and number of lasers that are phase locked}
In order to reduce the background intensities before calculating the IPR, we applied an intensity cutoff at $10\%$ (or $30\%$) of the maximal intensity for each far-field intensity distribution of the square (or random) array. The number of lasers that are phase locked (synchronized) $G$ was determined by dividing the square of the distance between nearby intensity peaks by the area of the central intensity peak \cite{Mahler19} and was set to zero below $M_{c}$ where sharp peaks do not exist. Specifically $G=g_{h}\times g_{v}$ (horizontally $g_{h}$ or vertically $g_{v}$) with $g=D/{\displaystyle \text{FWHM}}$ where $D$ is the distance between the central peak to the next peak and $\text{FWHM}$ is the full width half maximum of the central peak as shown in the cross section intensities of Fig.~\ref{fig:2_crwdsynch_square}b.

\subsection*{Determining the critical number of lasers $M_{c}$}
The critical number of lasers $M_{c}$ was determined by fitting the peak to background ratio ($PBR$) by the function $PBR=A+\frac{B}{1+exp(-k(M-M_{c}))}$, where $M$ is the number of lasers, $A$ and $B$ are the $PBR$ values of the low and high plateaus around the transition, $k$ is a constant that denotes the sharpness of the transition and $M_{c}$ is the critical number of lasers. From the fitting, the errors on the critical number of lasers $M_{c}$ was determined according to two-sigma ($95\%$) confidence.

\subsection*{Calculating the coupling $K$}
The coupling $K$ between the star and hub DCLs corresponds to the fraction of light transferred from the star lasers to the hub laser by the polarization beam splitter. This fraction is varied by rotating the quarter-wave plate in Fig.~\ref{fig:1_exp_skecth}(c). Using Malus’s law, the coupling can be calculated as $K=sin^{2}(4\theta)$, where $\theta$ is the 
angle between the horizontal axis and the fast-axis of the quarter-wave plate. 

\bibliography{biblio}



\section*{Acknowledgements}
The authors wish to acknowledge the Israel Science Foundation and the Israeli Planning and Budgeting Committee Fellowship Program for their support and thank Haim Nakav and Chene Tradonsky for valuable help. 





\end{document}